\documentclass[conference]{IEEEtran}
\IEEEoverridecommandlockouts
\usepackage{cite}
\usepackage{amsmath,amssymb,amsfonts}
\usepackage{algorithm}
\usepackage{algorithmic}
\usepackage{graphicx}
\usepackage{textcomp}
\usepackage{xcolor}
\usepackage{makecell}
\usepackage{multirow}

\usepackage[OT1]{fontenc}
\usepackage{caption}
\usepackage{booktabs} 
\def\BibTeX{{\rm B\kern-.05em{\sc i\kern-.025em b}\kern-.08em
    T\kern-.1667em\lower.7ex\hbox{E}\kern-.125emX}}

\begin{document}
\title{ATWM:Defense against adversarial malware based on adversarial training\\}

\author{\IEEEauthorblockN{1\textsuperscript{st}  Kun Li}
    \IEEEauthorblockA{
        \textit{PLA Information Engineering University}\\
        Zhengzhou, China \\
        moyue\_lk@foxmail}
    \and
    \IEEEauthorblockN{2\textsuperscript{nd}  Fan Zhang}
    \IEEEauthorblockA{
        \textit{PLA Information Engineering University}\\
        Zhengzhou, China \\
    }
    \and
    \IEEEauthorblockN{3\textsuperscript{rd}  Wei Guo}
    \IEEEauthorblockA{
        \textit{PLA Information Engineering University}\\
        Zhengzhou, China \\
    }
}

\maketitle

\begin{abstract}
    Deep learning technology has made great achievements in the field of image. In order to defend against malware attacks, researchers have proposed many Windows malware detection models based on deep learning. However, deep learning models are vulnerable to adversarial example attacks. Malware can generate adversarial malware with the same malicious function to attack the malware detection model and evade detection of the model. Currently, many adversarial defense studies have been proposed, but existing adversarial defense studies are based on image sample and cannot be directly applied to malware sample. Therefore, this paper proposes an adversarial malware defense method based on adversarial training. This method uses preprocessing to defend simple adversarial examples to reduce the difficulty of adversarial training. Moreover, this method improves the adversarial defense capability of the model through adversarial training. We experimented with three attack methods in two sets of datasets, and the results show that the method in this paper can improve the adversarial defense capability of the model without reducing the accuracy of the model.
\end{abstract}


\section{Introduction}
With the development of Internet technology, more and more researchers are focusing on the security of cyberspace. Malware is an important method of network attack, which poses a huge threat to the security of cyberspace. In recent years, deep learning has made great progress in image classification\cite{bochkovskiy2020yolov4}\cite{huang2017densely}, natural language processing\cite{devlin2018bert}\cite{vaswani2017attention}, speech signal processing\cite{sak2015fast}\cite{xiong2016achieving}, recommendation systems\cite{he2017neural}\cite{covington2016deep}, and machine translation\cite{sutskever2014sequence}, and researchers have begun combining deep learning techniques to detect malware. At present, the main research on malware detection based on deep learning is end-to-end detection. End-to-end malware detection only requires inputting the Portable Executable (PE) binary file into the detection model, and the model can output the detection result. Compared to traditional malware detection methods, deep learning malware detection automatically learns to sample features through neural networks, which reduces manual participation and reduces detection costs\cite{wang2021survey}. Deep learning has now achieved significant results in the field of end-to-end malware detection\cite{raff2017malware}\cite{tekerek2022novel}.

However, existing research shows that deep learning-based malware detection models are vulnerable to adversarial malware\cite{liu2019atmpa}\cite{mao2022adversarial}\cite{khormali2019copycat}\cite{demetrio2021functionality}. The adversarial attack can be divided into black box adversarial attack and white box adversarial attack according to whether the attacker can obtain the model weight, gradient, structure and other information. In the black box attack, the attacker cannot obtain important information such as model structure, model weight, model gradient, etc. The attacker can only obtain the input and output of the model, and obtain useful information from it. In the white box attack, the attacker can obtain important information such as model gradient, model structure, model weight parameters, etc., but the attacker cannot interfere with model training and can only implement the attack in the model testing stage.

To sum up, when applied, deep learning malware detection models are attacked by adversarial examples. Therefore, how to improve the adversarial defense capability of detection models has become a research topic of concern. The existing adversarial example defense research can be mainly divided into adversarial training\cite{madry2017towards}, randomization\cite{xie2017mitigating}, and denoising\cite{xu2017feature}, but the above research is based on the image domain. Malware is different from images. Small changes to the pixels of the image will not change its semantic information, but small changes to the content of the malware will cause it to lose malicious function and executable. Due to the difference between adversarial malware and adversarial image, the above defense methods cannot defend against adversarial malware and will affect the normal detection of malware detection models. Therefore, the existing adversarial example defense research cannot be directly applied to deep learning malware detection models. Therefore, defending against adversarial malware is very challenging. Our defense must reduce the negative impact of adversarial malware on the model, but not affect the normal detection of the malware detection model.

This paper proposes an adversarial malware defense method ATWM (Adversarial Training for Windows Malware) based on adversarial training. We evaluate the effectiveness of ATWM, and the results show that ATWM can improve the adversarial defense capability of malware detection models. The contributions of this paper are summarized as follows:
\begin{enumerate}
    \item We use preprocessing to filter low entropy perturbations to defend against BARAF attacks. Moreover, preprocessing reduces inefficient adversarial examples to make adversarial training less difficult.
    \item We use two perturbation generative approaches and two perturbation injection methods for stochastic combinations to generate diverse adversarial examples.
    \item We improve the robustness and adversarial defense of malware detection models through adversarial training.
\end{enumerate}

The remainder of the paper is organized as follows. In Section 2, we present work related to adversarial example defense. Section 3 describes the proposed ATWM. Section 4 reports the experimental results of ATWM. Finally, our paper is summarized in Section 6.
\section{Related work and background}
\subsection{Related work}
Adversarial examples have been widely proven to trick deep learning models into output errors. Therefore, researchers use it as an attack against deep learning models and propose various methods to defend against adversarial example attacks. Most of the existing adversarial example defense research focuses on image and text domains, so it cannot be directly applied to adversarial malware defense. However, the idea of adversarial example defense is universal. This section will introduce the existing image adversarial example defense ideas and research status. The existing adversarial defense ideas can be mainly divided into adversarial training, randomization and denoising.

\textbf{Adversarial Training:}Adversarial training is a general adversarial example defense method that adds adversarial example to training set of the model and enhances the robustness of the model by retraining to defend against adversarial example attacks. Goodfellow\cite{goodfellow2014explaining} et al. added adversarial examples generated by the Fast Gradient Sign Method (FGSM) to model training to enhance model robustness. After experimental evaluation, the model trained by this method can defend against the adversarial example generated by FGSM, but the model still shows very fragile when faced with more powerful adversarial attacks. Madry\cite{madry2017towards} et al. proposed to use Projected Gradient Descent (PGD) to generate adversarial examples for adversarial training to improve the robustness of the model. Experiments show that the robustness of the model trained by this method is greatly improved, and it can defend against typical adversarial attack methods such as FGSM, PGD, C\&W\cite{carlini2017towards}. However, this method consumes a lot of resources in adversarial training and cannot defend against other paradigms of adversarial attacks. The adversarial example is the key to adversarial training. To improve the quality of adversarial examples, ensemble adversarial training\cite{tramer2017ensemble} (EAT) is proposed. ETA adds adversarial examples from multiple pre-trained models to increase the diversity of adversarial examples. Experiments show that in some adversarial attacks, the adversarial defense capability of the EAT adversarial training model is better than that of the PGD adversarial training model.

In contrast to the above ideas, Lee\cite{lee2017generative} et al. proposed using Generative Adversarial Network\cite{goodfellow2020generative} (GAN) for adversarial training, rather than using a specific adversarial example generation algorithm. The GAN consists of the generator and discriminator. The method uses the generator to generate adversarial perturbations and adds perturbations to the original sample to generate adversarial examples. The discriminator model can be improved by feeding the original sample and adversarial example to the discriminator. Through the mutual adversarial training of the generator and discriminator, the generating perturbation ability of the generator can be improved. After the generative adversarial network is trained, use the adversarial example generated by it for adversarial training. This method can not only generate high-quality adversarial examples, but also train discriminators to detect adversarial examples, but GAN training is difficult.

\textbf{Randomization:}The aim of randomization is to reduce the attack capability of the adversarial example in order to improve the adversarial defense capability of the model. Xie\cite{xie2017mitigating} et al. use stochastic transform to input images to defend against adversarial examples. Such as stochastic image resizing, stochastic filling zeros on image edges. Dhillon\cite{dhillon2018stochastic} et al. proposed stochastic feature pruning to reduce the weight of adversarial perturbations on classification results. In addition to stochastic processing of input samples, Liu\cite{liu2018towards} et al. add stochastic noise to the network. This method achieves stochastic model inference by adding a noise layer in front of the convolutional layer of the model.

\textbf{Denoising:}Denoising reduces noise in samples to defend against adversarial attacks. The researchers found that perturbations added to images exhibit similar characteristics to noise, so noise reduction can be used to attenuate the attack power of adversarial examples. Xu\cite{xu2017feature} et al. proposed to reduce the high frequency noise in the input sample by reducing the bit depth of the image sample and blurring the image. Experiments show that this method can reduce the attack capability of the adversarial example and improve the adversarial defense capability of the model. In addition, some studies map the adversarial example to the original sample through generative adversarial network\cite{goodfellow2020generative} (GAN) and auto-encoder (AE). Input noise is reduced by the mapping process. The main work is Defense-GAN\cite{liu2018towards}, APE-GAN\cite{shen2017ape}, MagNet\cite{meng2017magnet}.

In conclusion, the above adversarial defense research cannot be applied directly to adversarial malware defense. Adversarial training is general, but methods that generate adversarial images cannot generate adversarial malware. Randomization and denoising will change the content of the PE file, resulting in a decrease in the accuracy of the malware detection model. Therefore, how to defend against adversarial malware based on adversarial training is the focus of this paper.

\subsection{Malware detection model}
\begin{figure*}[hbtp]
    \centering
    \includegraphics[width=\linewidth]{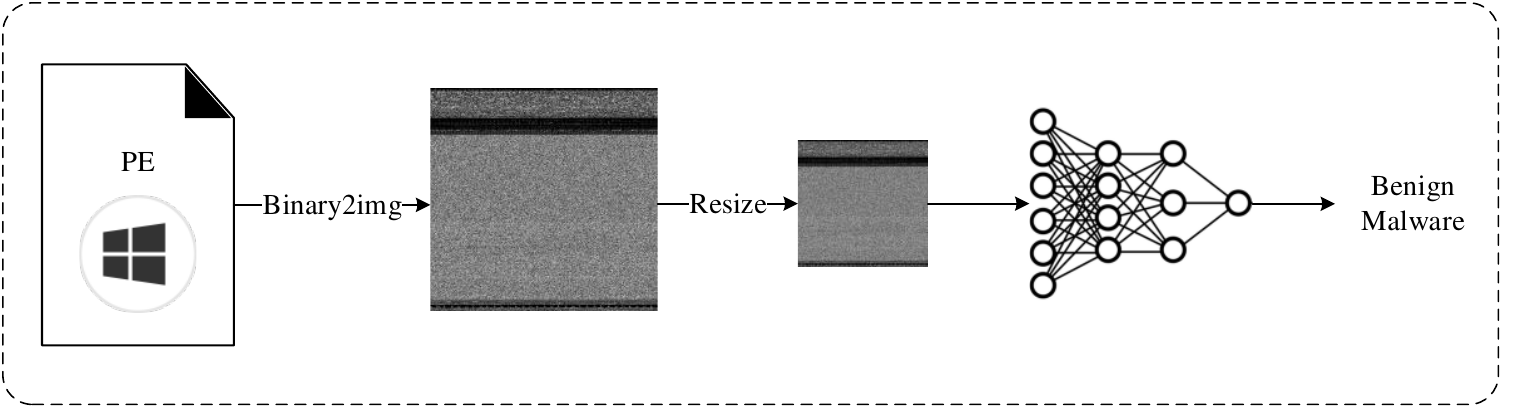}
    \caption{The byte-to-image malware detection model}
    \label{fig1}
\end{figure*}

The malware detection model in this paper is a byte-to-image malware detection model. Its process is shown in Figure\ref{fig1}, which is mainly divided into binary2img, resize, and classification\cite{cui2018detection}\cite{xiao2021image}\cite{tekerek2022novel}. The first is binary2img, which converts PE file binary bytes into gray images. The second is resizing, which resizes all gray images to the same size for parallel computing. Finally, classification, which feeds gray images into existing convolutional neural network models for classification. In summary, the input of this detection method is a gray image with size adjustment, so the input loses some features of the original data source, resulting in weak comprehensibility of the detection method. However, the size adjustment reduces resource consumption and improves detection speed.

In Binary2img, the specific steps follow the B2IMG algorithm\cite{tekerek2022novel}. First, the algorithm determines the gray image size according to the size of the PE file. The specific rules are that the gray image sizes of different PE files are different, but the width and height of the gray image are the same. That is, the width of the gray image multiplied by the height of the gray image is approximately equal to the size of the PE file. Secondly, the algorithm converts 8 unsigned binary numbers into decimal numbers between [0,255], and forms an image array according to the width and height of the gray image, and the insufficient area is supplemented by 0.

Currently, several models for detecting malware that converts bytes into images have been proposed. This paper selects the best performing model as the detection model of this paper\cite{tekerek2022novel}, and the convolutional neural network used is DenseNet121\cite{huang2017densely}.

\section{ATWM}
This section mainly introduces the specific details of the adversarial malware defense method based on adversarial training. Firstly, the overall process of ATWM is introduced, followed by the generation and injection perturbation methods used by ATWM, and finally the algorithm for generating various adversarial examples in ATWM is introduced.

\subsection{Problem formulation}
Deep learning can be divided into supervised machine learning and unsupervised learning according to different training methods. The current training method for deep learning is mainly supervised machine learning, and model $\mathcal{F}$ can be obtained through training. Model $\mathcal{F}$ is essentially a mapping relationship between the feature vector X and the label vector Y, where X and Y conform to the joint distribution P(X, Y). In supervised machine learning training, the loss function $l$ is defined, and the weight of the model is continuously adjusted with the goal of minimizing the loss function until the model learns the mapping relationship between X and Y. In training, the average loss function in the P distribution is the expected risk\cite{zhang2017mixup}, as shown in Equation\ref{eq1}. Therefore, the goal of supervised machine learning is to minimize the expected risk. However, in practical applications, it is impossible to obtain all the data that conform to the joint distribution P, and the distribution of the existing training data set can only be called empirical distribution. The average of the loss function in the empirical distribution $P_\sigma$ is called empirical risk\cite{zhang2017mixup}, as shown in Equation\ref{eq1}. When the difference between the empirical distribution $P_\sigma$ and the joint distribution P is very small, the trained model will be able to fully map the feature vector X to the label vector Y. However, when the data distribution $P_\sigma$ of the training set is quite different from the joint distribution P, the model obtained through empirical minimization training has poor robustness and will be vulnerable to adversarial examples.
\begin{equation}
    R(f)=\int l(f(x),y)dP(x,y)
    \label{eq1}
\end{equation}
\begin{equation}
    R_\sigma  (f)=\int l(f(x),y)dP_\sigma (x,y) =\frac{1}{n}\sum_{i=1}^{n}l(f(x_i ),y_i)
    \label{eq2}
\end{equation}

The essence of adversarial training is that by adding adversarial examples to the training data, the data distribution of the training set is similar to the joint distribution P. We can enhance the robustness of our malware detection model against adversarial example attacks by training it in the adversarial training set.

\subsection{ATWM framework}
\begin{figure*}[hbtp]
    \centering
    \includegraphics[width=\linewidth]{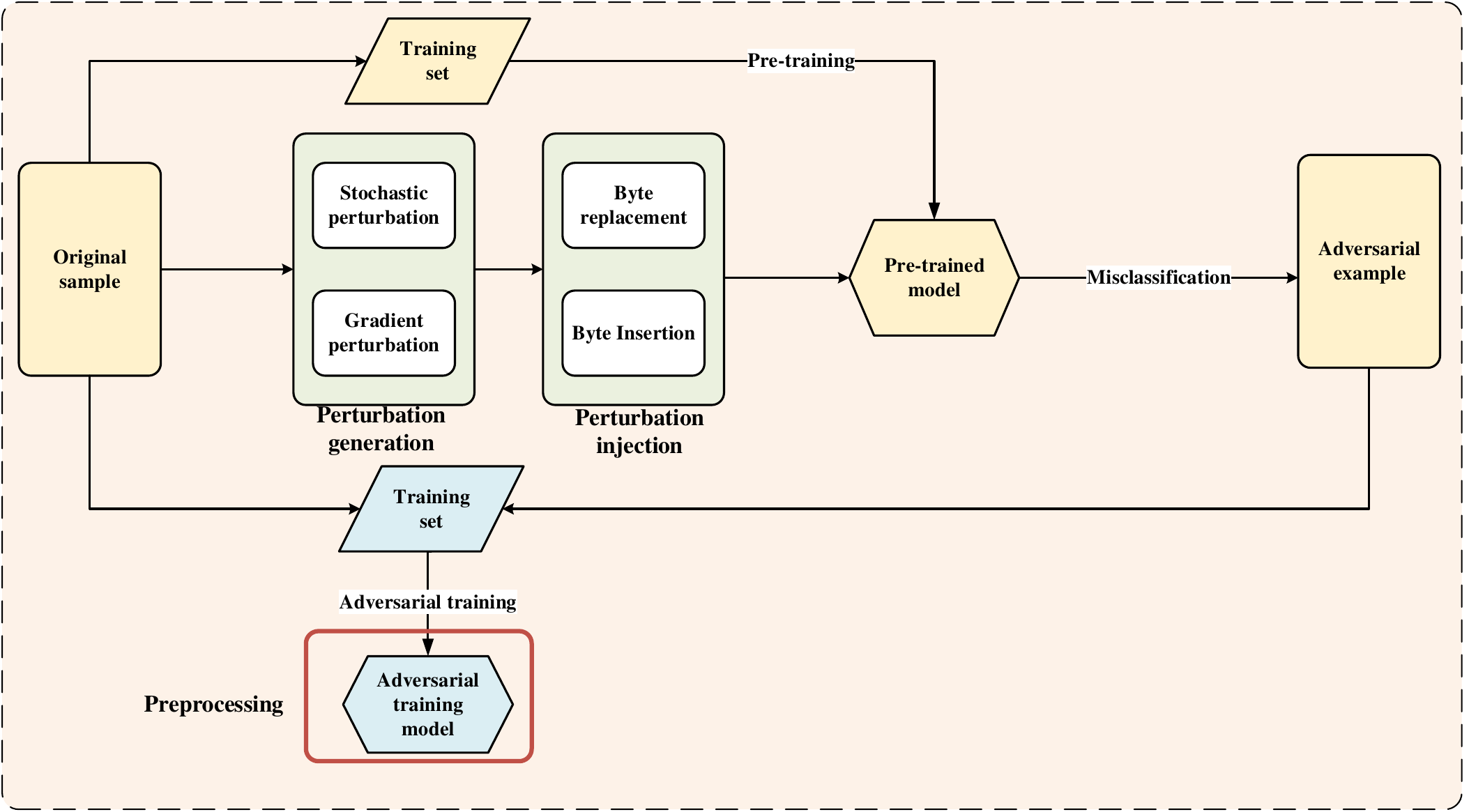}
    \caption{ATWM}
    \label{fig2}
\end{figure*}

The ATWM framework is shown in Figure\ref{fig2}, which is mainly divided into perturbation generation, perturbation injection, adversarial training, and preprocessing. First, the pre-trained model is obtained based on the original sample training set, and then perturbation is generated based on the pre-trained model, and perturbation is injected into the original sample to generate adversarial examples. When the pre-trained model misclassifies the adversarial example, the adversarial example is successfully generated. Finally, the original sample and adversarial example are combined into an adversarial training set, and the adversarial training set is used for training (adversarial training) to obtain a robustness-enhanced malware detection model. In addition, ATWM adds preprocessing before model input, which filters out simple perturbations with obvious features. Existing research has shown that adversarial training can improve the adversarial defense of the model but also reduce the accuracy of the model\cite{madry2017towards}. ATWM filters simple perturbations through preprocessing and reduces simple adversarial examples in the adversarial training dataset, which can reduce the decline in model accuracy.

\subsection{Preprocessing}
The preprocessing goal of ATWM is to filter perturbations with adversarial attack capability but obvious features. This type of perturbation can destroy the byte feature of PE files, and lower information entropy is the characteristic of this type of perturbation. In general, there is no content with very low information entropy in the PE file, and even if there is such a portion of the content, it is not used (such as bytes filling gaps). This type of adversarial example does not appear under normal circumstances and does not belong to the content of the joint distribution P. If such adversarial examples are added, the accuracy of the adversarial training model will be greatly reduced. Therefore, ATWM preprocesses PE files, removing the very low entropy content of the file information. The specific operation is shown in algorithm 1. The main principle of this algorithm is to remove sequences with low entropy in the byte sequence of PE files. First, obtain the byte sequence in the sliding interval and calculate the sequence entropy. When the entropy value is higher than the threshold, the sequence will be added to the preprocessed malware byte sequence, otherwise it will be discarded.
\begin{algorithm}[hbt]
    \caption{ATWM preprocessing}
    \begin{algorithmic}[1]
        \REQUIRE ~~ \\
        $x$, the original malware bytes; \\
        $\theta $, threshold; $l$, the sliding interval;\\
        \ENSURE ~~ \\
        $x'$,the malware bytes after preprocessing;
        \STATE $length=len(x)$
        \STATE $j \leftarrow 0, x' \leftarrow 0,i \leftarrow 0$
        \WHILE {$i<length$}
        \STATE $list_j=x_i$
        \STATE $j \leftarrow j+1, i \leftarrow i+1$
        \IF {$j>l$}
        \IF{$entropy(list)>\theta$}
        \STATE $x' \leftarrow x'+list$
        \ENDIF
        \STATE $j \leftarrow 0$
        \ENDIF
        \ENDWHILE
        \RETURN $x'$
    \end{algorithmic}
\end{algorithm}

\subsection{Perturbation generation and injection}
In this paper, various perturbation generation and perturbation injection methods are set up to generate diverse adversarial malware through stochastic combination methods. This section will introduce two perturbation generation methods and two perturbation injection methods.

\subsubsection{Perturbation generation}
The perturbations generated by ATWM are stochastic perturbation bytes and gradient perturbation bytes, which correspond to perturbations generated by different types of attacks. The content of stochastic perturbations is stochastic generated, as shown in Equation\ref{eq3}.
\begin{equation}
    s=random(0,255)
    \label{eq3}
\end{equation}
The gradient perturbation content is generated by the original PE file, and the generation method is shown in Equation\ref{eq4}, where $x$ is the byte of the original PE file, $g_sign$ is the gradient direction of the malware detection model, and $\eta$ is the perturbation intensity. The gradient perturbation content is obtained by adding the original PE file and the gradient perturbation, which retains the structure information of the original PE file. This type of perturbation is mainly generated by gradient-based attack methods, such as FGSM\cite{goodfellow2014explaining}, PGD\cite{madry2017towards}, etc.
\begin{equation}
    s=x+\eta g_{sign},\eta\in [0,255]
    \label{eq4}
\end{equation}

\subsubsection{Perturbation injection}
After obtaining the adversarial perturbation, ATWM injects the perturbation through byte replacement and byte insertion. Currently, adversarial examples of malware are presented as variants of malware samples. The byte replacement and byte insertion operations on this paper are equivalent to existing malware variant operations.

The first is byte replacement. The main characteristic of this type of operation is that part of the byte content in the PE file is replaced. When this part of the content has a significant impact on the model decision-making, replacing this part of the content will cause malware to evade detection of the model. The main representative operations are as follows:
\begin{enumerate}[noitemsep]
    \item Code Replacement: In the PE file, the original code is replaced with the same logical code, such as calculating constant values. The above operation will change part of the PE file, but in the PE file bytes it appears that part of the bytes will be replaced.
    \item Code Encryption: In a PE file, the content of the code is encrypted. For example, strings, text, etc. are encrypted. This operation changes part of the contents of the PE file, but in the PE file bytes, this operation replaces part of the bytes.
\end{enumerate}

The second is byte insertion. The main characteristic of this type of operation is that some content is inserted in the PE file. When this area has a significant impact on decision-making, inserting content near this area will cause the original features of the area to be destroyed, allowing malware to evade detection of the model. The main representative operations are as follows:
\begin{enumerate}[noitemsep]
    \item Junk code insertion: Junk code is inserted into the contents of the PE file, and the junk code does not interfere with the normal operation of the program. Such as empty NOP instructions, sequential addition and subtraction of register values, pushing values onto the stack and popping them up, computational constants, etc. There are various ways to implement the above operations in practice. But in PE file bytes, these operations will inject some bytes.
    \item Gap filling. Fill in any contents in the PE file byte gap. Due to the alignment mechanism of the PE file, there are many byte gaps in the PE file. The byte gap space is not used by the program and can be filled with any content, and the filling content does not affect the original function of the program. This operation will inject some bytes into the bytes of the PE file.
    \item Indirect call function: Call the function in the PE file by nesting multiple function calls, state synchronization calls, etc. This operation adds code logic to the code. But in the PE file bytes, the operation injects some bytes.
\end{enumerate}

The above operations have many ways in practice, but in PE file bytes, these operations add or replace some bytes. Both operations are equivalent to existing malware variant technology, but the content produced by the specific malware variant is unpredictable. Therefore, the main purpose of the above operation is to force the malware detection model to pay more attention to global features and reduce sensitivity to behaviors such as local content insertion and local content replacement, in order to improve the robustness and adversarial example defense capabilities of the model.

\subsection{Adversarial example generation}
The specific process of generating the adversarial example is shown in algorithm 2. We first obtain the length of the original malicious bytes, and combining the number of perturbation injections, stochastic generates a list of perturbation injection position coordinates (line1-line2). We sort the list of coordinates in reverse to avoid coordinate changes caused by byte insertion operations (line3). The injected perturbation has a stochastic value between 0 and the maximum perturbation (line4). We use two perturbation generation methods to generate perturbations and random selection of injection operations(line5-line6). We inject perturbations according to the coordinate list. Since the coordinate list is in reverse order, the perturbation of the injection does not change the subsequent injection coordinates. The original sample is injected multiple times to generate an adversarial example (line7-line11). Note that the adversarial malware generated in ATWM cannot be executed, but presents the same byte features as the executable adversarial example. Therefore, it can be used for adversarial training to improve the adversarial defense of the model.
\begin{algorithm}[hbt]
    \caption{ATWM generation adversarial example}
    \begin{algorithmic}[1]
        \REQUIRE ~~ \\
        $x$, the original malware bytes; \\
        $indexs $, the number of perturbations; \\
        $sizes$, the injection amount of perturbation;\\
        $\eta $,the intensity of the perturbation
        \ENSURE ~~ \\
        $x_{adv}$,the adversarial malware bytes;
        \STATE $length=len(x)$
        \STATE $list \leftarrow randlist(length,indexs)$
        \STATE $list \leftarrow sort(list)$
        \STATE $size \leftarrow random(0,sizes)$
        \STATE $s_1,s_2 \leftarrow generate(random(0,255),x+\eta g_{sign})$
        \STATE $operation \leftarrow random(replace , insert)$
        \STATE$x_1 \leftarrow x,x_2 \leftarrow x,i \leftarrow 0$
        \WHILE {$i<len(list)$}
        \STATE $index \leftarrow list_i,i \leftarrow i+1 $
        \STATE $x_1 \leftarrow operation(x_1,index,s_1),x_2 \leftarrow operation(x_2,index,s_2)$
        \ENDWHILE
        \STATE $x_{adv} \leftarrow x_1,x_2$
        \RETURN $x_{adv}$
    \end{algorithmic}
\end{algorithm}

\section{Experiment}
In order to verify the improved effect of ATWM on the robustness and adversarial defense capability of the model, the accuracy and adversarial defense capability of the model are evaluated. Existing studies have shown that adversarial training can improve the adversarial defense capability of the model, but reduce the detection accuracy of the model. Adversarial training should balance model accuracy with model adversarial defense capability. Therefore, this paper evaluates the effect of ATWM from two perspectives: model accuracy and adversarial defense capability. The experimental environment is based on Linux CentOS 7.9 operating system, equipped with Intel (R) Xeon (R) Gold 5218 CPU @2.30GHz (CPU), 256G (RAM), NVIDIA Tesla V100 32GB * 3 (GPU), based on PyTorch deep leaning framework, using Python to complete programming.

\subsection{Experimental datasets}
In this article, we use two datasets for experiments, the BIG2015 dataset and the VirusShare dataset. The BIG2015 data set comes from Microsoft\cite{ronen2018microsoft}. To ensure security, the PE file header has been removed and cannot be executed. The composition of the dataset is shown in Table\ref{tab1}. There is a large imbalance in the number of categories in the BIG2015 dataset, resulting in insufficient robustness of the trained model.
\begin{table}[!htb]
    \centering
    \resizebox{.75\columnwidth}{!}{
        \begin{tabular}{lll}
            \toprule
            Class & Name           & Amount \\
            \midrule
            0     & Ramnit         & 1533   \\
            1     & Lollipop       & 2478   \\
            2     & Kelihosver3    & 2942   \\
            3     & Vundo          & 474    \\
            4     & Simda          & 42     \\
            5     & Tracur         & 751    \\
            6     & Kelihosver1    & 398    \\
            7     & Obfuscator.ACY & 1228   \\
            8     & Gatak          & 1013   \\
            9     & Total          & 10859  \\
            \bottomrule
        \end{tabular}}
    \caption{\label{tab1}BIG2015 dataset}
\end{table}
The VirusShare dataset is a non-public dataset that is collected by itself and comes from the VirusShare website\footnote{https://virusshare.com/} and Windows system. The PE file in the dataset can be executed because it has not been processed. The composition of its data set is shown in Table\ref{tab2}. Among them, Benign comes from the Windows system, and Malware comes from the VirusShare website.
\begin{table}[!htb]
    \centering
    \resizebox{.6\columnwidth}{!}{
        \begin{tabular}{lll}
            \toprule
            Class & Name    & Amount \\
            \midrule
            0     & Benign  & 10613  \\
            1     & Malware & 10308  \\
            2     & Total   & 20921  \\
            \bottomrule
        \end{tabular}}
    \caption{\label{tab2}VirusShare dataset}
\end{table}
The above data set is divided into training set, validation set, and test set according to the ratio of 6:2:2. During the experiment, the malware detection model is trained through the training set, and the model that performs best on the validation set is selected as the final model, and the accuracy of the model are verified on the test set. It is worth noting that the adversarial attack process is time-consuming, so 500 samples were stochastic selected from the test set as the adversarial defense capability test data set.

\subsection{ATWM parameters and examples}
The ATWM parameter settings are shown in Table\ref{tab3}. The preprocessing threshold is set to 1, and when the perturbation content is a single byte, the entropy value is 0. When bytes fluctuate in the range of difference 2, the entropy approximation is 1. Therefore, ATWM will filter byte sequences in PE files where byte fluctuations do not exceed 2.
\begin{table}[!htb]
    \centering
    \resizebox{.95\columnwidth}{!}{
        \begin{tabular}{lll}
            \toprule
            Name                              & Parameter & Value \\
            \midrule
            Threshold                         & $\theta$  & 1     \\
            Sliding interval                  & $l$       & 10 KB \\
            Number of Perturbations           & $indexs$  & 2     \\
            Maximum injection amount          & $sizes$   & 0.2   \\
            The intensity of the perturbation & $\eta $   & 0.3   \\
            \bottomrule
        \end{tabular}}
    \caption{\label{tab3}ATWM parameters}
\end{table}
The gray image of ATWM-generated adversarial malware is shown in Figure 2. The adversarial malware cannot be executed, but has the same gray image characteristics as the executable adversarial malware. The example diagram shows that the adversarial malware retains some features of the original sample, but parts of the adversarial malware have been inserted or replaced with adversarial perturbations.
\begin{figure*}[hbtp]
    \centering
    \includegraphics[width=\linewidth]{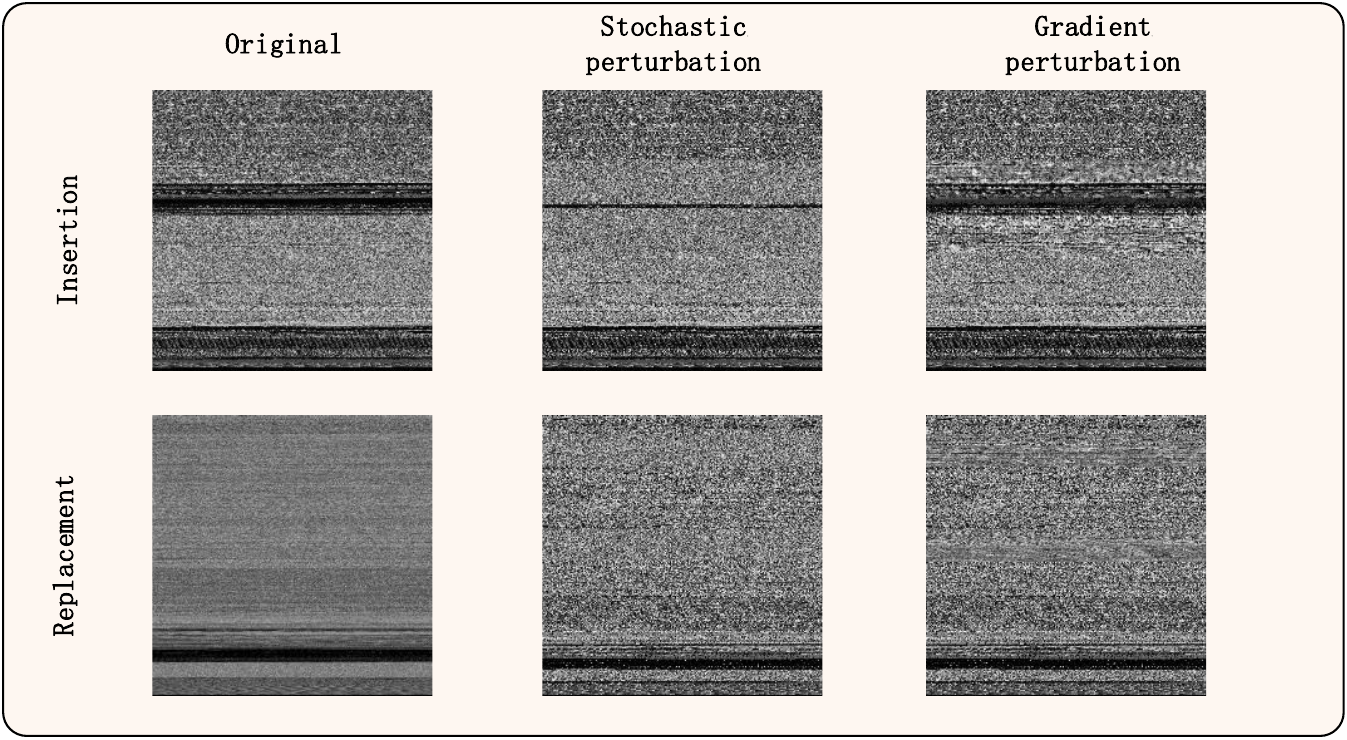}
    \caption{Example of adversarial malware}
    \label{fig3}
\end{figure*}

\subsection{Model accuracy}
ATWM adds preprocessing before malware detection models, and uses a diverse adversarial malware data set for adversarial training. This section evaluates model accuracy on two datasets. In both datasets, SGD dynamic optimizer is used for training. The parameters of the optimizer are: learning rate is 0.1, decay rate is 0.6, step size is 5, and the number of training iterations is 100 rounds. First, the accuracy of the original model (OM), preprocessing + original model (P + OM), and preprocessing + adversarial training model (P + ATM) were tested on the VirusShare dataset. The experimental results are shown in Table\ref{tab4}. The results show that the adversarial training model obtained by ATWM does not have the problem of precision degradation on the VirusShare dataset.
\begin{table}[!htb]
    \centering
    \resizebox{.85\columnwidth}{!}{
        \begin{tabular}{llll}
            \toprule
            Evaluation & OM     & P + OM & P + ATM \\
            \midrule
            accuracy   & 0.9142 & 0.9116 & 0.9140  \\
            auc        & 0.9145 & 0.9115 & 0.9138  \\
            macro avg  & 0.9142 & 0.9115 & 0.9140  \\
            \bottomrule
        \end{tabular}}
    \caption{\label{tab4}Model accuracy (VirusShare dataset)}
\end{table}

Secondly, the accuracy of the original model, preprocessing + original model, and preprocessing + adversarial training model were tested on the BIG2015 dataset. The experimental results are shown in Table\ref{tab5}. Experimental results show that the adversarial training model obtained by ATWM does not decrease the accuracy of the BIG2015 dataset, but improves the accuracy of the model. The main reason is that preprocessing filters out meaningless bytes in the original sample, making the features of the original sample more prominent.
\begin{table}[!htb]
    \centering
    \resizebox{.85\columnwidth}{!}{
        \begin{tabular}{llll}
            \toprule
            Evaluation & OM     & P + OM & P + ATM \\
            \midrule
            accuracy   & 0.9655 & 0.9793 & 0.9720  \\
            auc        & 0.9729 & 0.9750 & 0.9705  \\
            macro avg  & 0.9389 & 0.9538 & 0.9405  \\
            \bottomrule
        \end{tabular}}
    \caption{\label{tab5}Model accuracy (BIG2015 dataset)}
\end{table}

\subsection{Model adversarial defense capabilities}
To evaluate the effectiveness of ATWM, the adversarial defense capabilities of the model were tested using existing adversarial attack methods. The attack methods include black box attack methods and white box attack methods. Black box attacks are GAMMA\cite{demetrio2021functionality} and BARAF\cite{mao2022adversarial}, and white box attacks is COPYCAT\cite{khormali2019copycat}. Note that for uniform comparison, the amount of perturbation injection is used as the attack strength control variable.

The first is GAMMA. GAMMA is a black box attack method based on genetic algorithms. This method generates adversarial malware through genetic algorithms to attack malware detection models. The GAMMA parameters are set as follows: the number of genetic algorithm iterations is set to 20, the population size is set to 50, the crossover parameter is set to 0.7, and the mutation parameter is set to 0.8. The experimental results are shown in Table\ref{tab6}. Experimental results show that the adversarial defense capability of the model obtained by the ATWM method is better than that of the original model on both datasets. In addition, as the attack intensity increases, the model accuracy gradually decreases, but the ATWM method is still effective.
\begin{table}[!htb]
    \centering
    \resizebox{.9\columnwidth}{!}{
        \begin{tabular}{lllll}
            \toprule
            Dataset    & Model   & 0.1        & 0.2        & 0.3        \\
            \midrule
            BIG2015    & OM      & 0.664      & 0.624      & 0.606      \\
            BIG2015    & P + ATM & \bf{0.878} & \bf{0.824} & \bf{0.772} \\
            VirusShare & OM      & 0.72       & 0.654      & 0.62       \\
            VirusShare & P + ATM & \bf{0.772} & \bf{0.71}  & \bf{0.662} \\
            \bottomrule
        \end{tabular}}
    \caption{\label{tab6}GAMMA attack results}
\end{table}

The second is BARAF, which is a black box attack method based on image texture feature destruction, which generates adversarial malware by destroying the texture feature of the original sample to attack the malware detection model. In the experiment, FF byte with the best attack effect on the original paper was used as the perturbation byte. Experimental attack results are shown in Table\ref{tab7}. The experimental results show that the perturbation used by BARAF is a simple perturbation, which can be completely filtered by ATWM preprocessing, and will not lead to a decrease in the accuracy of the model. In addition, as the amount of perturbation increases, the defense effect does not decrease.
\begin{table}[!htb]
    \centering
    \resizebox{.9\columnwidth}{!}{
        \begin{tabular}{lllll}
            \toprule
            Dataset    & Model   & 0.1        & 0.2        & 0.3        \\
            \midrule
            BIG2015    & OM      & 0.51       & 0.484      & 0.436      \\
            BIG2015    & P + ATM & \bf{0.97}  & \bf{0.968} & \bf{0.968} \\
            VirusShare & OM      & 0.868      & 0.82       & 0.816      \\
            VirusShare & P + ATM & \bf{0.913} & \bf{0.912} & \bf{0.912} \\
            \bottomrule
        \end{tabular}}
    \caption{\label{tab7}BARAF attack results}
\end{table}

Finally, there is COPYCAT, a white box attack method based on model gradients that generates adversarial malware by injecting adversarial perturbations into the end of the original sample. During the attack, the perturbation strength of COPYCAT is set to 0.1, and perturbation is generated for the entire original sample, and perturbation near the injection area is selected as injection perturbation. The results of the attack experiment are shown in Table\ref{tab8}. The results show that ATWM improves the defense capability of the model against COPYCAT attacks. As the amount of perturbation increases, the defense capability does not decrease significantly.
\begin{table}[!htb]
    \centering
    \resizebox{.9\columnwidth}{!}{
        \begin{tabular}{lllll}
            \toprule
            Dataset    & Model   & 0.1         & 0.2        & 0.3        \\
            \midrule
            BIG2015    & OM      & 0.744       & 0.762      & 0.748      \\
            BIG2015    & P + ATM & \bf{ 0.914} & \bf{0.90}  & \bf{0.804} \\
            VirusShare & OM      & 0.868       & 0.802      & 0.806      \\
            VirusShare & P + ATM & \bf{0.89}   & \bf{0.854} & \bf{0.816} \\
            \bottomrule
        \end{tabular}}
    \caption{\label{tab8}COPYCAT attack results}
\end{table}

\section{Conclusion}
The deep learning malware end-to-end detection model is vulnerable to adversarial attack, but existing adversarial defense research cannot be applied directly to adversarial malware defense. Aiming at this problem, this chapter proposes an adversarial malware defense method (ATWM) based on adversarial training. ATWM filters out simple perturbations through preprocessing, and generates adversarial examples through various stochastic combinations for adversarial training to improve the adversarial defense ability of the model. Experimental evaluation on two sets of different types of data sets shows that ATWM can improve the adversarial defense ability of the model without causing a significant decrease in model accuracy. ATWM is highly scalable. In future work, the effect of ATWM can be improved by adding more adversarial malware generative approaches to stochastic combinations. In addition, ATWM is versatile, and in the next work, ATWM can be applied to multiple malware detection models.

\bibliography{sample}

\end{document}